\documentclass[sigconf]{acmart}

\usepackage{multirow}
\usepackage{booktabs} 
\usepackage{amsmath}
\usepackage{graphicx}
\usepackage[belowskip=-12pt]{caption}
\usepackage{rotating}
\usepackage{tabularx}
\usepackage{subfig}
\usepackage{wrapfig}
\usepackage{listings}
\usepackage{color, colortbl}
\usepackage{balance} 
\usepackage{lscape}
\usepackage{hyperref}
\usepackage{xspace}
\usepackage{framed}
\usepackage{syntax}
\usepackage[framemethod=TikZ]{mdframed}
\usepackage{soul}
\usepackage{flushend}

\definecolor{codegreen}{rgb}{0,0.6,0}
\definecolor{codegray}{rgb}{0.5,0.5,0.5}
\definecolor{codepurple}{rgb}{0.58,0,0.82}
\definecolor{backcolour}{rgb}{0.95,0.95,0.92}
\definecolor{Gray}{gray}{0.1}
\lstdefinestyle{mystyle}{
	backgroundcolor=\color{backcolour},   
	commentstyle=\color{codegreen},
	keywordstyle=\color{magenta},
	numberstyle=\tiny\color{codegray},
	stringstyle=\color{codepurple},
	basicstyle=\scriptsize,
	breakatwhitespace=false,         
	breaklines=true,                 
	captionpos=b,                    
	keepspaces=true,                 
	numbers=left,                    
	numbersep=5pt,                  
	showspaces=false,                
	showstringspaces=false,
	showtabs=false,                  
	tabsize=2
}

\graphicspath{{./figures/}}
\newcommand{\figref}[1]{Fig.~\ref{#1}}

\newcommand{\fignref}[1]{Figure~\ref{#1}}

\newcommand{\secref}[1]{\S\ref{#1}}

\newcommand{\commitcount}{100\xspace}
\newcommand{\totalcommit}{500\xspace}
\newcommand{\etal}{{\em et al.}\xspace}

\newcommand{\sof}{\textit{Stack Overflow}\xspace}
\newcommand{\caffe}{\textit{Caffe}\xspace}

\newcommand{\keras}{\textit{Keras}\xspace}

\newcommand{\tensor}{\textit{Tensorflow}\xspace}
\newcommand{\theano}{\textit{Theano}\xspace}
\newcommand{\torch}{\textit{Torch}\xspace}

\newcommand{\gh}{\textit{Github}\xspace}

\newcounter{NumObservations}
\stepcounter{NumObservations}
\definecolor{shadecolor}{rgb}{.9,.9,.9}

\newcommand{\finding}[1]{
	
	\begin{shaded}
		\vspace{-.85em}
		\textit{{\bf{Finding \arabic{NumObservations}}}: #1.}\\
		\vspace{-1.75em}
	\end{shaded}
	\stepcounter{NumObservations}	
}
\setcopyright{rightsretained}

\acmDOI{10.475/123_4}

\acmISBN{123-4567-24-567/08/06}

\acmConference[ESEC/FSE 2019]{The 27th ACM Joint European Software Engineering Conference and Symposium on the Foundations of Software Engineering}{26--30 August, 2019}{Tallinn, Estonia}

\acmYear{2019}
\copyrightyear{2019}

\acmArticle{4}
\acmPrice{15.00}

\begin{document}
	\title{A Comprehensive Study on Deep Learning Bug Characteristics}
	
\author{Md Johirul Islam}
\email{mislam@iastate.edu}
\affiliation{%
	\institution{Iowa State University}
	\streetaddress{216 Atanasoff Hall}
	\city{Ames}
	\state{IA}
	\postcode{50010}
}

\author{Giang Nguyen}
\email{gnguyen@iastate.edu}
\affiliation{%
	\institution{Iowa State University}
	\streetaddress{216 Atanasoff Hall}
	\city{Ames}
	\state{IA}
	\postcode{50010}
}
\author{Rangeet Pan}
\email{rangeet@iastate.edu}
\affiliation{%
	\institution{Iowa State University}
	\streetaddress{216 Atanasoff Hall}
	\city{Ames}
	\state{IA}
	\postcode{50010}
}

\author{Hridesh Rajan}
\email{hridesh@iastate.edu}
\affiliation{%
	\institution{Iowa State University}
	\streetaddress{216 Atanasoff Hall}
	\city{Ames}
	\state{IA}
	\postcode{50010}
}

	\begin{abstract}
Deep learning has gained substantial popularity in recent years.
Developers mainly rely on libraries and tools to add deep learning capabilities to their software.
What kinds of bugs are frequently found in such software?
What are the root causes of such bugs?
What impacts do such bugs have?
Which stages of deep learning pipeline are more bug prone?
Are there any antipatterns?
Understanding such characteristics of bugs in deep learning software has 
the potential to foster the development of better deep learning 
platforms, debugging mechanisms, development practices, and encourage the 
development of analysis and verification frameworks. 
Therefore, we study 2716 high-quality posts from \sof and 500 bug fix commits from \gh about five popular deep learning libraries \caffe, \keras, \tensor, \theano, and \torch to understand the types of bugs, root causes of bugs, impacts of bugs, bug-prone stage of deep learning pipeline as well as whether there are some common antipatterns found in this buggy software. 
The key findings of our study include: data bug and logic bug are the most severe bug types in deep learning software appearing more than 48\% of the times, major root causes of these bugs are Incorrect Model Parameter (IPS) and Structural Inefficiency (SI) showing up more than 43\% of the times.
We have also found that the bugs in the usage of deep learning libraries have some common antipatterns that lead to a strong correlation of bug types among the libraries. 
\end{abstract}
	
%
%
%

	\maketitle
	
	\section{Introduction}
\label{sec:intro}

A class of machine learning algorithms known as {\em deep learning} has
received much attention in both academia and industry.
These algorithms use multiple layers of transformation functions to convert
input to output, each layer learning successively higher-level of abstractions 
in the data. The availability of large datasets has made it feasible to train
(adjust the weights of) these multiple layers.
While the jury is still out on the impact of deep learning on overall 
understanding of software's behavior, a significant uptick in its usage
and applications in wide ranging areas combine to warrant research on 
software engineering practices in the presence of deep learning.  
This work focuses on the characteristics of bugs in software that makes 
use of deep learning libraries.

Previous work on this topic generally fall under two categories: 
those that have studied bugs in the implementation of machine learning 
libraries themselves, and those that have studied bugs in the usage of a 
specific deep learning library. 
A key work in the first category is Thung \etal \cite{thung2012empirical} 
who studied bugs in the implementation of three machine learning systems 
Mahout, Lucene, and OpenNLP.
In the second category, Zhang \etal \cite{zhang2018empirical} have studied
bugs in software that make use of the \tensor library.
While both categories of approaches have advanced our knowledge of ML 
systems, we do not yet have a comprehensive understanding of bugs encountered 
by the class of deep learning libraries. 

This work presents a comprehensive study of bugs in the usage of deep learning  
libraries. We have selected top five popular deep learning libraries
\caffe \cite{jia2014caffe},  \keras \cite{chollet2015keras}, \tensor \cite{abadi2016tensorflow}, \theano \cite{team2016theano},  and \torch \cite{collobert2002torch} based on the user counts from
developers Q\&A forum \sof.
While each of these libraries are for deep learning they have different design goals.
For example, \tensor focuses on providing low-level, highly configurable facilities 
whereas \keras aims to provide high-level abstractions hiding the low-level details.
\theano and \torch are focused on easing the use of GPU computing to make deep learning
more scalable. 
Thus, studying them simultaneously allows us to compare and contrast their design goals
{\em vis-\`a-vis} bugs in their usage.

We have used two sources of data in our study: posts about these libraries on \sof and
also \gh bug fix commits.
The first dataset gives us insights into bugs that developers encounter when building
software with deep learning libraries. 
A number of these bugs would, hopefully, be fixed based on the discussion in Q\&A forum.
The second dataset gives us insights into bugs that were found and fixed in open 
source software. Our study focuses on following research questions and compares
our findings across the five subject libraries.

\noindent\textbf{RQ1: (Bug Type)} What type of bugs are more frequent? 

\noindent\textbf{RQ2: (Root cause)} What are the root causes of bugs?

\noindent\textbf{RQ3: (Bug Impact)} What are the frequent impacts of bugs?

\noindent\textbf{RQ4: (Bug prone stages)} Which deep learning pipeline stages are more vulnerable
to bugs? 

\noindent\textbf{RQ5: (Commonality)} Do the bugs follow a common pattern?

\noindent\textbf{RQ6: (Bug evolution)} How did the bug pattern change over time?

{\bf Findings-at-a-glance.\ }
Our study show that most of the deep learning bugs are {\em Data Bugs} and 
{\em Logic Bugs}~\cite{beizer1984software}, 
the primary root causes that cause the bugs are
Structural Inefficiency (SI) and Incorrect Model Parameter (IPS)~\cite{beizer1984software}, most of the
bugs happen in the Data Preparation stage of the deep learning pipeline. 
Our study also confirms some of the findings of \tensor conducted by Zhang \etal
\cite{zhang2018empirical}.
We have also studied some antipatterns in the bugs to find whether there is 
any commonality in the code patterns that results in bugs. 
Our findings show that there is strong correlation among the distribution of
bugs as well as in the antipatterns.
Finally, we conclude with a discussion on our findings suggesting immediate actions 
and future research directions based on these findings. 


	\section{Methodology}
\label{sec:method}

\begin{table}[t]
	\centering
	\caption{Summary of the dataset used in the Study}
	\setlength{\tabcolsep}{4.8pt}
	\renewcommand{\arraystretch}{0.5}
	\begin{tabular}{lcccc}
\hline
\multicolumn{1}{|c|}{\multirow{2}{*}{Library}} & \multicolumn{2}{c|}{\sof}                          & \multicolumn{2}{c|}{\gh}                                    \\ \cline{2-5} 
\multicolumn{1}{|c|}{}                         & \multicolumn{1}{c|}{\# Posts} & \multicolumn{1}{c|}{\# Bugs} & \multicolumn{1}{c|}{\# Commits} & \multicolumn{1}{c|}{\# Bugs} \\ \hline
\multicolumn{1}{|l|}{\caffe}                    & \multicolumn{1}{r|}{183}      & \multicolumn{1}{r|}{35}      & \multicolumn{1}{r|}{100}        & \multicolumn{1}{r|}{26}      \\ \hline
\multicolumn{1}{|l|}{\keras}                    & \multicolumn{1}{r|}{567}      & \multicolumn{1}{r|}{162}     & \multicolumn{1}{r|}{100}        & \multicolumn{1}{r|}{348}     \\ \hline
\multicolumn{1}{|l|}{\tensor}               & \multicolumn{1}{r|}{1558}     & \multicolumn{1}{r|}{166}     & \multicolumn{1}{r|}{100}        & \multicolumn{1}{r|}{100}     \\ \hline
\multicolumn{1}{|l|}{\theano}                   & \multicolumn{1}{r|}{231}      & \multicolumn{1}{r|}{27}      & \multicolumn{1}{r|}{100}        & \multicolumn{1}{r|}{35}      \\ \hline
\multicolumn{1}{|l|}{\torch}                    & \multicolumn{1}{r|}{177}      & \multicolumn{1}{r|}{25}      & \multicolumn{1}{r|}{100}        & \multicolumn{1}{r|}{46}      \\ \hline
\multicolumn{1}{|l|}{Total}                    & \multicolumn{1}{r|}{2716}     & \multicolumn{1}{r|}{415}     & \multicolumn{1}{r|}{500}        & \multicolumn{1}{r|}{555}     \\ \hline
                                               & \multicolumn{1}{l}{}          & \multicolumn{1}{l}{}         & \multicolumn{1}{l}{}            & \multicolumn{1}{l}{}        
\end{tabular}
	\label{tbl:dataset}
\end{table}
\subsection{Data Collection}
In our study two different data sources are used. \sof posts and \gh bug fix commits are the sources of data we used for studying the bugs in deep learning software. A summary of the dataset is shown in Table \ref{tbl:dataset}.
\subsubsection{\sof Data Collection}
To study bugs in deep learning software, we have collected data from \sof, a well-known Q\&A site for developers to discuss
software development problems. 
The data collection process consists of two steps. 

In the first step, we select candidate posts discussing deep learning libraries. 
We focus on five deep learning libraries: \caffe, \keras, \tensor, \theano, and \torch. 
These are the five most discussed deep learning libraries on
\sof. 
We did that by searching for posts tagged with \caffe, \keras, \tensor, \theano, and \torch. 
When posts are about specific libraries, 
they are more likely to talk about bugs in using deep learning libraries. 
Using these criteria, we selected all posts about these five libraries. 
We further filtered the posts that did not contain any source code
because posts about bugs usually contain code snippets. Moreover, 
we reduced the number of posts by selecting the posts whose scores, 
computed as the difference between the number of its upvotes and the 
number of its downvotes, were greater than 5 to focus on
the high-quality posts and keep the manual effort manageable. 
After this step, in total, we selected 183, 567, 1558, 231, and 177  
posts for \caffe, \keras, \tensor, \theano, and \torch,   respectively.

In the second step, we manually read these candidates to identify the ones about
bugs. After that, the second and the third authors manually reviewed the candidates. For 
each post, we read the question and all answers focusing on the best-accepted 
one. If the best-accepted answer was to fix the usages of the deep learning API(s) in 
the question, we considered that post as talking about deep learning bugs. After 
this step, we found 35, 162, 166,  27, and 25
bugs for \caffe, \keras, \tensor, \theano, and \torch respectively.

\subsubsection{\gh Data Collection}
\label{subsec:gh-data}

\gh is a large source of deep learning repositories. We mine the \gh commits to 
study the change in the commits and to check and confirm the bug patterns that we studied 
from \sof.
The data collection process consists of two steps. 

First, we collect all the repositories of \caffe, \keras, \tensor, \theano, and \torch.
After that, we mine all the commits whose title contain word \texttt{"fix"} of these libraries. Then, 
we randomly select \commitcount commits for each libraries from mined commits and classify them.

Secondly, we use the same process that we used for \sof. Specifically, the second and the third authors 
manually studied the \totalcommit commits and separately label them. After that, these two authors compare 
their results to fix the conflict in the labeling process. We study every line of change in each commits; therefore, some commits have more than one bugs and some commit does not have bug. Overall, we got 26, 348, 100, 35, and, 46  bugs for the commits of \caffe, \keras, \tensor, \theano, and \torch, respectively.
\subsection{Classification}
\label{subsec:classification}

In our classification, we focus on three criteria which are bug types, root causes and effects of bug.
The classification scheme used for labeling of the bugs in each of these three criteria discussed in \secref{subsec:type-bugs}, \secref{subsec:classify-root}, and \secref{subsec:classify-effects}.
We have also classified the bugs into different deep learning stages \cite{stages}.

To label the bug types we followed the classification from an already
existing well vetted taxonomy \cite{beizer1984software} and appended on top of that. The added types were based on the data that we studied following an open coding scheme. 

The bugs may have different root causes and effects. A supervised pilot study and open coding schemes were used to identify the effects that are possible through these bugs. We have adapted the classification scheme of root causes and bug effects from \cite{zhang2018empirical} and added on top of that as found from the study of the posts. 
The third author studied the posts initially to finalize the classification scheme for bug types, root causes and effects. We followed the open coding scheme and pilot study was conducted to get agreement on the classification.  

We also classified the bugs into different stages of the pipeline to understand which stages are more vulnerable to bugs. Deep learning process can be divided into seven stage pipeline \cite{stages}.
The stages are data collection, data preparation, choice of model, training,
evaluation, hyper parameter tuning and prediction. Among the seven stages, the first
one is not related to software development. The other stages are related to software development,
and are supported by the deep learning libraries through their APIs. 
We use these stages to label the bugs into different stages. 

\subsection{Labeling the Bugs}
\label{subsec:label-bug}

Once we have all the classification criteria, we used those criteria to label
the posts. The second and the third authors independently studied the posts. We measured the inter rater aggrement among the labellers using Cohen's Kappa coefficient \cite{viera2005understanding} when 5\%, 10\%, 20\% , 30\%, 40\%, 50\%,
60\%, 70\%, 80\%, 90\% and 100\% of the posts were labeled. 
After 5\% labeling, the Cohen's Kappa coefficient was close to 0. Then we
conducted a training session among the raters to clarify the labeling and
what they mean. 
After the training session, we conducted another pilot study at 10\% including
the first 5\%. This time the Cohen's Kappa coefficient was 82\%. We again discussed the
results and find out the reasons for major disagreements. We then discussed those
cases further through examples and continued labeling. 
The Cohen's Kappa coefficient was more than 90\% in subsequent pilot studies.  

The labeling effort was continuously being monitored with the help of Kappa
coefficient to understand the agreement. We conducted reconciling efforts ideally at
every 10\% interval of the labeling. The posts where there was disagreement
between the raters were further discussed in the presence of a supervisor. After
discussion and arguments a common label was given. Finally all the bugs were
given a common label.

\subsection{Types of Bugs in Deep Learning Software}
\label{subsec:type-bugs}

Developers often confront different types of bugs while trying to write deep learning software. To
understand those bugs and their root causes, we have classified them into
different categories. The classification has been done on the basis of all the
\sof posts that we analyzed. The classification is adapted from \cite{beizer1984software}, where a well organized taxonomy of bugs is presented. 



\subsubsection{API Bug}
This group of bugs is caused by deep learning API. Generally, when a developer uses a deep learning API, different bugs associated with that API are inherited automatically without the knowledge of the user. The prime causes for triggering of deep learning API bugs can be because of the change of API definition with different versions, lack of inter-API compatibility and sometimes wrong or confused documentation. 
\subsubsection{Coding Bug}
This kind of bugs originate due to mistakes in coding syntax. This in turn, introduces other types of bugs in the software which lead to either run time error or incorrect results. A big percentage of the deep learning bugs that we have checked arises from syntactic mistakes or a certain scenario which cannot be fixed by changing only some lines of code; hence, needs to change the whole module. Though a robust compiler usually takes care of the basic coding mistakes, in certain scenarios, this type of bugs are not captured by the compiler resulting in wrong output. 
\subsubsection{Data Bug}
This bug may arise if an input to the deep learning software is not properly formatted or cleaned well before processing in any deep learning model. This type of bug occurs before data enters into the deep learning model. It is not because of the wrong deep learning model, rather it is purely based on the type and structure training or test data.  Similar to coding bugs, data bugs are usually flagged by the compiler, but in some scenarios, it can pass unchecked through the compilation process and generate erroneous results. 

\subsubsection{Structural Bug(SB)}
A vast majority of the deep learning bugs are occurring due to incorrect definitions of the deep learning model's structure. These include mismatch of dimensions between different
layers of deep learning models, the presence of anomaly between the training and test
datasets, use of incorrect data structures in implementing a particular
function, etc. This type of bugs can be further classified into another four
categories.
\paragraph{Control and Sequence Bug}
This subclass of the bug is caused by the wrong structure of control flow. In many scenarios, due to wrong if-else or loop guarding condition, the model does not perform as expected. This type of bug either leads to a crash when a part of deep learning model does not work or, leads to incorrect functionality due to mishandling of data through the layers.

\paragraph{Data Flow Bug}
The main difference between the Data Flow Bug and the Data Bug is the place of origination. If a bug occurs due to the type or shape mismatch of input data after it has been feed to the deep learning model, it will be called Data Flow Bug. It includes the scenarios when model layers are not in synchronization because of different data shape used in consecutive layers. To fix these bugs, developers need to modify the model or reshape the data. 
\paragraph{Initialization Bug}
In deep learning, Initialization Bug means the parameters or the functions are not initialized properly before they are used. This type of bugs would not necessarily produce run time error but it will simply make the model perform worse. Here, the definition of functions includes both user-defined and API defined. We also categorize a bug into this category when the API has not been initialized properly.
\paragraph{Logic Bug}
In deep learning, the logical understanding of each stage of the pipeline is an integral part of the coding process. With an incorrect logical structure of the deep learning model, the output of a program may result in either a runtime error or a faulty outcome. These bugs are often generated in the absence of proper guarding conditions in the code or trying to implement a feature which is not possible in the given structure of the deep learning model. 

\paragraph{Processing Bug}
One of the most important decisions in the deep learning model structure is to choose the correct algorithm for the learning process. In fact, different deep learning algorithms can lead to different performances and outputs \cite{gomez2016empirical}. Also, to make different layers be compatible with each other, the data types of each layer need to follow a contract between them. Processing Bugs happen due to the violation of these contracts and wrong choice of algorithms.

\subsubsection{Non Model Structural Bug(NMSB)}
Unlike SB, NMSB is created outside the modeling stage. In other words, this bug can happen in any deep learning stage except the modeling stage such as the training stage or the prediction stage. 
NMSB has similar subcategories with SB. Subcategories of NMSB are Control and Sequence Bug, Logic Bug, Processing Bug, and Initialization Bug. 
We do not define Non Model Structural Data Flow Bug like  Structural Data Flow Bug because Data Bug already covers the meaning of Non Model Structural Data Flow Bug.

\paragraph{Control and Sequence Bug}
This subclass is similar to Control and Sequence Bug in SB. The bug is caused by an incorrect structure of control flow like wrong if-else condition; however, this kind of bug happens outside modeling stage.
\paragraph{Initialization Bug}
This subclass is similar to Initialization Bug in SB. The bug is caused by initializing a parameter or a function in a wrong way; however, this kind of bug happens outside modeling stage.
\paragraph{Logic Bug}
This subclass is similar to Logic Bug in SB. The bug is caused by misunderstanding how case statements and logical operators behave singly; however, this kind of bug happens outside modeling stage.
\paragraph{Processing Bug}
This subclass is similar to Processing Bug in SB. The bug is caused by an incorrect choice of algorithm; however, this kind of bug happens outside modeling stage.

\subsection{Classification of Root Causes of bugs}
\label{subsec:classify-root}

\subsubsection{Absence of inter API compatibility.}
The main reason for these bugs is the inconsistency of the combination of two different kinds of
libraries. For example, a user cannot directly use \texttt{Numpy} function in \keras because
neither \tensor backend nor \theano backend of \keras has the implementation of
\texttt{Numpy} function. 

\subsubsection{Absence of type checking.}
The major effect of the bugs is crash. 
This kind of bugs involves a type mismatch
problem when calling API methods. These bugs are usually mistakes
related to the use of wrong type of parameters in an API.

\subsubsection{API Change.}
The reason for these bugs is the release of
the new version of a deep learning library. In other words, the bug happens when the new
API version is not backward compatible with its previous version. 
For example, a user updates the new version of a deep learning library which has new API
syntax; however, the user does not modify his/her code to fit with the new
version, which leads to the API change bug.

\subsubsection{API Misuse.}
This kind of bugs
often arises when users use a deep learning API without fully understanding. 
Missing conditions can be one kind of API misuse, and this bug occurs when a
usage does not follow the API usage constraints to ensure certain required
conditions. Crash is the main effect of these bugs.

\subsubsection{Confusion with Computation Model.} 
These bugs happen when a user gets confused about the function of deep learning API, which leads
to the misuse of the computation model assumed by the deep learning library.
For instance, a user gets confused between the graph construction and the evaluation
phase. 

\subsubsection{Incorrect Model Parameter or Structure (IPS)}
IPS causes problems with constructing the deep learning model, e.g. incorrect model structures
or using inappropriate parameters. 
IPS is a common bug in the deep learning software because of both the lack of deep learning
knowledge among the users and the incomprehension of deep learning models. 
This kind of bugs causes the functional incorrectness; thus, the effect of this
bug is crash.

\subsubsection{Others.}
These bugs are not related to deep learninng software. In other words, these bugs are mostly related to
mistakes in the development process like incorrect syntax. 

\subsubsection{Structure Inefficiency (SI)}
SI causes problems related to modeling stage in deep learning software like IPS; however, SI leads to bad performance of the deep learning software while IPS leads to
crash.

\subsubsection{Unaligned Tensor (UT)}
These bugs often occur in the computation graph construction phase.
When a user builds the computation graph in deep learning process, they have to provide
correct input data with required specifications to a deep learning API; 
however, many users do not know exactly their specifications, or they misunderstand
API signature, which leads to UT bugs.

\subsubsection{Wrong Documentation.}
Incorrect information in library documentation leads to these bugs. Deep learning library
users may face this kind of bugs when they read an incorrect definition or
an incorrect usage of a deep learning API from documentation.

\subsection{Classification of Effects of bugs}
\label{subsec:classify-effects}

\subsubsection{Bad performance.}
Bad performance or poor performance is one of common kind of effect in deep learning
software. Furthermore, the major root causes of this effect are SI or CCM 
that are related to model
construction. Even though developers can use the deep learning libraries correctly, they still face deep learning model construction problems because APIs in these libraries are highly abstract. 

\subsubsection{Crash.} Crash is the most frequent effect in deep learning. 
In fact, any kind of bugs can lead to Crash. A symptom of crash is that the
software stops running and prints out an error message.

\subsubsection{Data Corruption.} Data corruption happens when the data is
corrupted as the data flows through the network. This effect is a consequence of misunderstanding the
deep learning algorithms or APIs. When Data Corruption occurs, a user will receive unexpected outputs.

\subsubsection{Hang.} Hang effect is caused when a deep learning software ceases to respond to inputs.
Either slow hardware or inappropriate deep learning algorithm can lead to Hang. A symptom of Hang 
is that the software runs for a long period of time without providing the desired output.

\subsubsection{Incorrect Functionality}
This effect occurs when the software behaves in an unexpeced way without any runtime or compile-time error/warning. 
This includes the incorrect output format,
model layers are working desirably, etc.

\subsubsection{Memory out of bound}
Deep learning software often halts due to unavailability of the memory resources. This can
be caused by, either the wrong model structure or, not having enough computing
resources to train a particular model.
	\section{Frequent bug types}
In this section, we explore the answer to \textbf{RQ1} through statistically analyzing the labeled
data. 
The normalized distribution of bug types in \sof data is shown in \fignref{fig:bugtype}. The distribution of bugs shown in \fignref{fig:bugtype}  and the \sof and \gh data in Table \ref{tbl:gitsobugs} shows the presence of different kinds of bugs in both \sof and \gh for the deep learning libraries we have studied. We present some of the key findings related to bug types in the following subsections. 
\subsection{Data Bugs}
\begin{figure}
	\includegraphics[keepaspectratio = True, scale = .25]{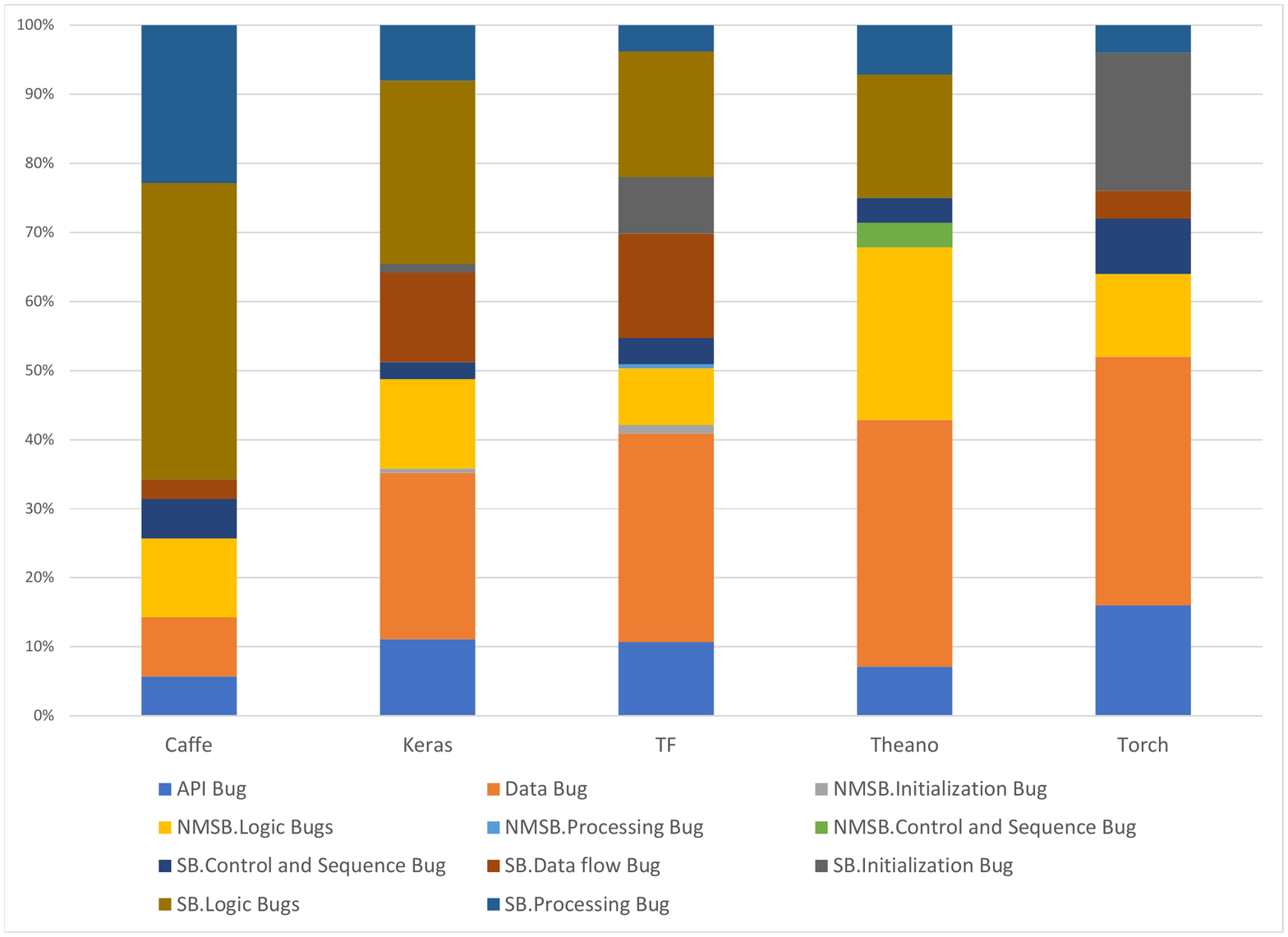}
	\vspace{-2.8em}
	\caption{Distribution of Bug Types in \sof}
	\label{fig:bugtype}
\end{figure}
\finding{Data Bugs appear more than 26\% of the times}

From \fignref{fig:bugtype} we see that among the bug types the Data Bugs appear most of the time (26\%) in all the libraries.  
In the studied \sof data, we have seen 30\% of the posts in \tensor, 24\% posts
in \keras, 36\% posts in \torch, 35\% posts in \theano, and 9\% posts in \caffe have
Data Bugs.
Data bugs mostly appear due to the absence of data pre-processing like feature
engineering, data validation, data shuffling, etc. 

The large percentage of Data Bugs indicate data pre-processing related difficulties are appearing quite often which
could be addressed by some data verification tools. If we can provide some
static analysis tools using modern abstract data types like \textit{DataFrame}
and the properties of the model, that would help the deep learning community. 
For example, a developer is trying to read some image files using the following
method\footnote{\url{https://tinyurl.com/y3v9o7pu}}.

\begin{lstlisting}
def _read32(bytestream):
	dt = numpy.dtype(numpy.uint32).newbyteorder('>')
	return numpy.frombuffer(bytestream.read(4), dtype=dt)
\end{lstlisting}
The developer eventually got stuck with the following error while trying to
train the model using the data returned by the previous library call.
\begin{lstlisting}[language = python]
TypeError: only integer scalar arrays can be converted to a scalar index   
\end{lstlisting}
An expert suggested an answer to change the last return statement with the
following, which solved the problem and was accepted by the developer:
\begin{lstlisting}[language = python]
return numpy.frombuffer(bytestream.read(4), dtype=dt)[0]
\end{lstlisting}
The bug is hard to fix by just looking at the error message. It is difficult to
identify the exact reason of bug which led the developer to post a question on 
\sof and the question was upvoted by other
fellow developers as a qualified post.

\subsection{Structural Logic Bugs}

\finding{\caffe has 43\% Structural Logic Bugs}
The second major bug type is Structural Logic Bug in \sof which was expected from our
initial hypothesis based on a pilot study. 
\caffe has more Structural Logic Bugs in \sof compared to other libraries. 
This indicates that the majority of the bugs in \caffe are made during
construction and logical organization of the model. Other libraries also have significant portion of Structural Logic Bugs ranging from 0\% - 27\%.
\subsection{API Bugs}
\finding{\torch, \keras, \tensor have 16\%, 11\% and 11\% API bugs respectively}
In deep learning libraries API changes sometimes break the entire production
code. The implicit dependence between libraries cause problems when
one library has some major changes. For example, when \texttt{Numpy} is updated \tensor,
\keras software may fail. \keras often uses \tensor or \theano as backend and hence update of \tensor or \theano can cause the software developed using \keras to crash.
API bugs are seen to appear more often in  \keras and \tensor as shown in
\fignref{fig:bugtype}. More than 81\% of the API bugs are from \keras
and \tensor.
For example, in the following code snippet extracted from \sof we see a scenario
where the developer trying to train a model fails due to the upgrade of APIs
and changing the keyword names in the API signature of \keras. 

\begin{table}
	\centering
	\caption{Statistics of Bug Types in \sof and \gh}
	\label{tbl:gitsobugs}
	\setlength{\tabcolsep}{0.5pt}
	\footnotesize
	\renewcommand{\arraystretch}{0.5}
	\begin{tabular}{|l|c|c|c|c|c|c|c|c|c|c|c|}
		\hline
		\multirow{2}{*}{}             		& \multicolumn{2}{c|}{Caffe} & \multicolumn{2}{c|}{Keras} 	& \multicolumn{2}{c|}{TF} 	& \multicolumn{2}{c|}{Theano}	& \multicolumn{2}{c|}{Torch} 	 		 		 	& \multirow{2}{*}{P value}\\ \cline{2-11} 
		& \rotatebox[origin=c]{90} {SO}  & \rotatebox[origin=c]{90} {GitHub}  & \rotatebox[origin=c]{90} {SO}   & \rotatebox[origin=c]{90} {GitHub}        & \rotatebox[origin=c]{90} {SO}       &\rotatebox[origin=c]{90} {GitHub}        & \rotatebox[origin=c]{90} {SO}       &
		\rotatebox[origin=c]{90} {GitHub}       & \rotatebox[origin=c]{90} {SO}       &
		\rotatebox[origin=c]{90} {GitHub}       & \\ \hline
		API Bug                       		 & 6\%  & 0\%		& 11\% & 57\%		& 11\% & 72\%     & 7\%     & 3\%		& 16\% & 2\%                    &0.3207 \\ \hline
		Data Bug                      		 & 9\%  & 49\%		& 24\% & 8\%		& 30\% & 0\%       & 35\%   & 17\% 		& 36\%  & 15\%                  &0.3901   \\ \hline
		NMSB.Control and Sequence Bug & 0\%  & 8\%		& 0\%  & 0\%		& 0\%  & 0\%      & 4\%    & 0\%		& 0\%     & 7\%                   &0.3056 \\ \hline		
		NMSB.Initialization Bug       		& 0\%  & 0\%		& 1\%  & 0\%		& 1\%  & 0\%      & 0\%    & 3\% 		& 0\%    & 0\%                    &0.7655  \\ \hline
		NMSB.Logic Bugs               		& 11\% & 0\% 		& 13\% & 2\%		& 8\%  & 0\%      & 25\%   & 6\% 		& 12\%   & 7\%                  & 0.0109  \\ \hline
		NMSB.Processing Bug           	& 0\%  & 0\%		& 0\%  & 0\%		& 1\%  & 0\%      & 0\%    & 3\%		& 0\%     & 7\%                     &0.2323   \\ \hline
		SB.Control and Sequence Bug   	& 6\%  & 12\%		& 2\%  & 0\%		& 4\%  & 0\%      & 4\%    & 3\%		& 8 \%    & 9\%                     &1.0000  \\ \hline
		SB.Data flow Bug              		& 3\%  & 8\%		& 13\% & 26\%		& 15\% & 0\%      & 0\%    & 14\%		& 4\%    & 16\%                   &0.2873 \\ \hline
		SB.Initialization Bug         		& 0\%  & 0\%		& 1\%  & 0\%		& 8\%  & 1\%      & 0\%    & 23\%		& 20\%   & 11\%                   &0.8446  \\ \hline
		SB.Logic Bugs                	 	& 42\% & 15\%		& 27\% & 3\% 		& 18\% & 23\%     & 18\%   & 14\%		& 0\%    & 13\%                 &0.3442 \\ \hline
		SB.Processing Bug             		& 23\% & 8\%		& 8\%  & 4\% 		& 4\%  & 4\%      & 7\%    & 14\%		& 4\%     & 13\%                     &0.8535  \\ \hline
	\end{tabular}
\vspace{-.85em}
\end{table}

\begin{lstlisting}
model.fit(trainX, trainY, epochs=100, batch_size=1, verbose=2)
\end{lstlisting}
The developer will get the error because \texttt{epochs} keyword does not exist in
version 2+ of \keras.
\begin{lstlisting}[language = python]
model.fit(trainX, trainY, batch_size=1, verbose=2, epochs = 100) File
"/usr/local/lib/python2.7/site-packages/keras/models.py", line 612, in fit
str(kwargs)) Exception: Received unknown keyword arguments: {'epochs': 100}
\end{lstlisting}
To fix this error, the developer needs to change from \texttt{epochs} to \texttt{nb_epoch}
\begin{lstlisting}[language = python]
model.fit(trainX, trainY, nb_epoch=100, batch_size=1, verbose=2)
\end{lstlisting}

\subsection{Bugs in \gh projects}
We have also analyzed the distributions of bugs in some \gh bug fix commits. The
distribution of bugs across different libraries in \gh data is shown in Table
\ref{tbl:gitsobugs}. We computed the P value using t-test where one distribution
is bug type in \gh for all the libraries and the other distribution is bug type
for all the libraries in \sof. 
\finding{All the bug types have a similar pattern in \gh and \sof for all the libraries}
We analyze the \sof and \gh result using the t-test to find whether the
distributions differ significantly. We use 95\% significant level to find the
difference beween \sof and \gh results for each of the bug type
In our analysis the null hypothesis is:
\textbf{\textit{H$_{0}$: The distributions are same}}.
If we fail to reject this null hypothesis using the t-test then we can say the
distributions follow the same pattern in both \sof and \gh data.

We see that for all the bug types except Non Model Structural Logic Bug
the P value is greater than 5\% indicating they have a similar pattern as we fail to reject the null hypothesis.

	\section{Root Cause}
\begin{figure}
	\includegraphics[keepaspectratio = True, scale = .25]{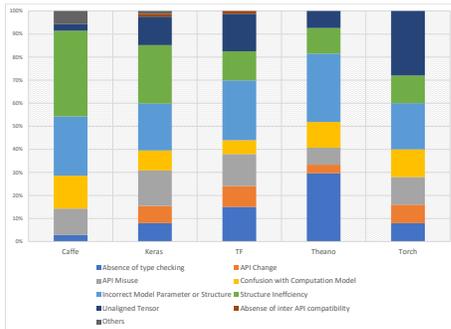}
	\vspace{-2.8em}
	\caption{Stack Overflow Root Cause Classification}
	\label{fig:rootcause}
\end{figure}

In this section, we present the analyses and findings to answer \textbf{RQ2} identifying major root causes of bugs in deep learning software.  
The normalized distribution of root causes in \sof code snippets is shown in \fignref{fig:rootcause}. The data in Table \ref{tbl:rootcausetbl} shows the presence of different categories of root causes in both \sof and \gh for the deep learning libraries and presents P value showing the similarity of distributions using t-test. 
We discuss the significant root causes in the following subsections.
\subsection{Incorrect Model Parameter (IPS)}

\finding{IPS is the most malicious root cause resulting in average 24\% of the bugs across the libraries}
IPS results in bugs that causes the program to crash at runtime and the execution does not succeed. 
In \tensor and \theano IPS leads other root causes in causing bugs having 26\% and 26\%  of the total share of root causes, respectively. 

\subsection{Structural Inefficiency (SI)}
\finding{\keras, \caffe have  25\% and 37\% bugs that are resulted from SI}
SI bugs do not cause the program to crash. These bugs often yield suboptimal performance of the deep learning model. These bugs have more relation to QoS or non-functional requirements. 
For example, a programmer is trying to train a model to recognize handwritten digits but the accuracy does not improve and stays constant from epochs  2 - 10. \footnote{\url{https://stackoverflow.com/questions/37213388/keras-accuracy-does-not-change}}
\begin{lstlisting}
Epoch 1/10
2394/2394 [==============================] - 0s - loss: 0.6898 - acc: 0.5455 - val_loss: 0.6835 - val_acc: 0.5716
Epoch 2/10
2394/2394 [==============================] - 0s - loss: 0.6879 - acc: 0.5522 - val_loss: 0.6901 - val_acc: 0.5716
.........
Epoch 10/10
2394/2394 [==============================] - 0s - loss: 0.6877 - acc: 0.5522 - val_loss: 0.6849 - val_acc: 0.5716
1027/1027 [==============================] - 0s
\end{lstlisting}

The problem that was pointed out by an expert, which solved the performance degradation bug is following:
\begin{lstlisting}
#In summary, replace this line:
model.compile(loss = "categorical_crossentropy", optimizer = "adam")
#with this:
from keras.optimizers import SGD
opt = SGD(lr=0.01)
model.compile(loss = "categorical_crossentropy", optimizer = opt)
\end{lstlisting}
The answer suggested to change \texttt{optimizer} for enhancing the performance.
\begin{table}[]
	\centering
	\caption{ Statistics of the Root Causes of Bugs}
	\label{tbl:rootcausetbl}
	\setlength{\tabcolsep}{0.05pt}
	\footnotesize
	\renewcommand{\arraystretch}{0.5}
	\begin{tabular}{|l|c|c|c|c|c|c|c|c|c|c|c|}
		\hline
		\multirow{2}{*}{}            &\multicolumn{2}{c|}{Caffe}	 &\multicolumn{2}{c|}{Keras}	& \multicolumn{2}{c|}{TF} 		& \multicolumn{2}{c|}{Theano}	&\multicolumn{2}{c|}{Torch} 		 		 		  & \multirow{2}{*}{P value}\\ \cline{2-11} 
		& \rotatebox[origin=c]{90} {SO}       &
		\rotatebox[origin=c]{90} {GitHub}       & \rotatebox[origin=c]{90} {SO}       &
		\rotatebox[origin=c]{90} {GitHub}        & \rotatebox[origin=c]{90} {SO}       &
		\rotatebox[origin=c]{90} {GitHub}        & \rotatebox[origin=c]{90} {SO}       &
		\rotatebox[origin=c]{90} {GitHub}       & \rotatebox[origin=c]{90} {SO}       &
		\rotatebox[origin=c]{90} {GitHub}        &\\ \hline
		
		Absense of inter API compatibility     		& 0\%     & 0\%		& 1\%     & 0\%		& 1\%  & 0\%      	&0\%      & 0\%			& 0\%     & 0\%     		      		      		      &0.1411 \\ \hline
		Absence of type checking               		& 3\%     & 12\%	& 8\%     & 3\%		& 15\% & 15\%     	&30\%    & 20\%		& 8\%     & 13\%     		    		      		     &0.9717  \\ \hline
		API Change                             			& 0\%     & 0\%		& 7\%     & 51\%	& 9\%  & 58\%    	&4\%      & 0\%			& 8\%     & 2\%      		      		     		      &0.2485 \\ \hline
		API Misuse                             			& 11\%    & 0\%	& 15\%    & 4\%	& 14\% & 0\%      	&7\%      & 3\%			& 12\%    & 2\%      		      		      		       &0.0003  \\ \hline
		Confusion with Computation Model     		& 14\%    & 28\%	& 9\%     & 1\%		& 6\%  & 10\%     	&11\%     & 3\%		& 12\%    & 4\%      		      		      		     &0.7839  \\ \hline
		Incorrect Model Parameter or Structure 	& 26\%    & 31\%	& 21\%    & 30\%	& 26\% & 16\%     	&30\%     & 14\%		& 20\%    & 19\%     		      		     		     &0.5040  \\ \hline
		Others                                 			& 0\%     & 0\%		& 0\%     & 0\%		& 0\%  & 0\%      	&0\%      & 0\%			& 0\%     & 2\%      		      		      		      &0.3466 \\ \hline
		Structure Ineffciency                  			& 37\%    & 12\%	& 26\%    & 5\%	& 13\% & 1\%      	&11\%     & 26\%		& 12\%    & 38\%     		     		      		     &0.7170 \\ \hline
		Unaligned Tensor                       		& 3\%     & 19\%	& 12\%    & 5\%	& 16\% & 0\%      	&7\%      & 34\%		& 28\%    & 20\%     		     		      		     &0.7541 \\ \hline
		Wrong Documentation                    		& 6\%     & 0\%		& 1\%     & 1\%		& 0\%  & 0\%      	&0\%      & 0\%			& 0\%     & 0\%      		      		      		      &0.3402\\ \hline
		
	\end{tabular}
\end{table}
\subsection{Unaligned Tensor (UT)}
\finding{\torch has 28\% of the bugs due to UT}
In deep learning, tensor dimensions are important for successful construction of the model. 
\tensor, \keras, \torch, \theano, \caffe have respectively 16\%, 12\%, 28\%, 7\% and 3\% of bugs due to UT respectively. In \torch UT is the highest root cause of bugs. 
\subsection{Absence of Type checking}
\finding{\theano has 30\% of the bugs due to the absence of type checking}
Most of the deep learning libraries are written in Python. Due to the dynamic nature of Python, the problem of the absence of type checking is felt strongly in these libraries. The absence of type checking leads to 30\% of the bugs in \theano, 8\% of the bugs in \keras and 15\% of the bugs in \tensor. 

\subsection{API Change}
\finding{\tensor and \keras have 9\% and 7\%  bugs due to API change}
In deep learning libraries, API change tends to have a drastic effect. These libraries are interdependent. So, API change in one library breaks other libraries.  

\subsection{Root Causes in \gh data}
\finding{Except API Misuse all other root causes have similar patterns in both \gh and \sof root causes of bugs}
We computed the P value at 95\% significant level for both the \sof and \gh data for all the root causes in the five libraries. We see that, P value for API Misuse root cause is much less than 5\% indicating API Misuse in \sof and \gh has different distribution compared to other root causes as we reject the null hypothesis. The other root causes are similar for both \sof and \gh data as their P value is greater than 5\%.

\subsection{Relation of Root Cause with Bug Type}
\finding{SI contributes  3\% - 52\% and IPS contirbutes  24\% - 62\% of the bugs related to model}
\begin{figure}
	\includegraphics[keepaspectratio = True, scale= .3]{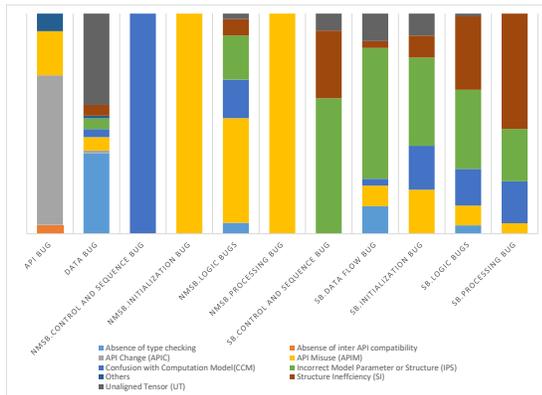}
	\caption{Relation between Root Causes  and Types of Bugs }
	\label{fig:bugcause}
\end{figure}

We have seen from \fignref{fig:bugcause} that most of the non model related bugs are caused by API Misuse (6\% - 100\%). Non Model Structural Initialization Bugs and Non Model Structural Processing Bugs are caused by API Misuse in 100\% of the time in our studied data. 
Interestingly in API Bug API Change plays the vital role (68\%) compared to API Misuse (20\%); however, the model related bugs are more vulnerable to IPS and SI root causes. We see from \fignref{fig:bugcause} that  Structural Control and Sequence Bug, Structaral Data Flow Bug, Structural Initialization Bug, Structural Logic Bug, Structural Processing Bug which are related to model are caused by SI 31\%, 3\%, 10\%, 33\% and 53\%  of the times respectively and caused by IPS 62\%, 59\%, 40\%, 36\%, 24\% of the times respectively.

	\section{Impacts from Bugs}
\begin{figure}
	\includegraphics[keepaspectratio = True, scale= .2]{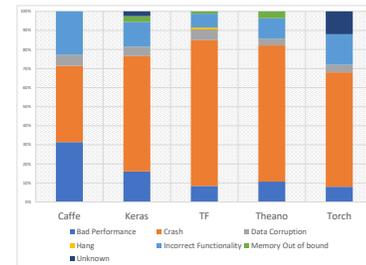}
	\vspace{-2.8em}
	\caption{Distribution of Bug Effects in \sof}
	\label{fig:impact}
\end{figure}

In this section, we explore the answer to \textbf{RQ3} to understand the major effects of bugs in deep learning software. 
The normalized distribution of effects of \sof is shown in \figref{fig:impact}. The data in Table \ref{tbl:impval} shows the presence of different kinds of effects in both \sof and \gh for the deep learning libraries. 
We discuss some of the major effects of bugs in deep learning software in the rest of this section. 

\subsection{Crash}
\finding{In average more than 66\% of the bugs cause crash of the programs}
Our analysis reveals that, the most severe effect of bugs is Crash. 
In deep learning, the bugs mostly cause total failure of the program. 
In all the libraries Crash is the top impact ranging from 40\% - 77\% as shown in \fignref{fig:impact}.

\subsection{Bad Performance}
\finding{In \caffe, \keras, \tensor, \theano, \torch  31\%, 16\%, 8\%, 11\%, and 8\% bugs caused bad performance respectively}
Bad performance is often a concern for deep learning software developers. Even though the model trains successfully, during the evaluation or prediction phase the model may give very poor accuracy in classifying the target classes. 

For example, in the following code snippet the user had low accuracy after traning because of the use of incorrect value of parameter \texttt{nb_words}. The user should use \texttt{nb_words + 1} instead of \texttt{nb_words} as answered by an expert. \footnote{\url{https://stackoverflow.com/questions/37817588/masking-for-keras-blstm}}
\begin{lstlisting}
embedded = Embedding(nb_words, output_dim=hidden, input_length=maxlen)(sequence)
\end{lstlisting}

\subsection{Incorrect Functionality}
\finding{12\% of the bugs in average in the libraries cause Incorrect Functionality }
Incorrect functionality happens when the behavior of the software reflects some unexplainable outcome which is not expected from the logical organization of the model or from previous experience of the developer. 

For example, in the following code snippet the user wants to convert the image to a $28*28$ \texttt{Numpy} array; however, the output is a black image.\footnote{\url{https://stackoverflow.com/questions/42353676/display-mnist-image-using-matplotlib}}
\begin{lstlisting}
with tf.Session() as sess:
    first_image = mnist.train.images[0]
    first_image = np.array(first_image, dtype='uint8')
    pixels = first_image.reshape((28, 28))
    plt.imshow(pixels, cmap='gray')
\end{lstlisting}
The user got incorrect output because of casting a float array to uint8, which will convert all the pixels to 0 if they are less than 1. To fix the problem, the user can multiply the array with 255 as suggested by an answer.
\theano has a higher percentage of posts about incorrect functionality problems more common than bad performance.

\subsection{Effects of Bugs in \gh}
\finding{For all the libraries the P value for \sof and \gh bug effects reject the null hypothesis to confirm that the bugs have similar effects from \sof as well as \gh bugs}
The P value is shown in Table \ref{tbl:impval} shows that Bad Performance in \sof and \gh have 79\% of P value which indicates that they are very similar. Crash has P value of 50\% in \sof and \gh indicating they also can not reject the null hypothesis with strong confidence. None of the impacts reject the null hypothesis at 95\% significance level. 
\begin{table}[]
\centering
\caption{Effects of Bugs in \sof and \gh}
\label{tbl:impval}
\setlength{\tabcolsep}{1.5pt}
\footnotesize
\renewcommand{\arraystretch}{0.5}
\begin{tabular}{|l|c|c|c|c|c|c|c|c|c|c|c|}
\hline
\multirow{2}{*}{}             & \multicolumn{2}{c|}{Caffe}	& \multicolumn{2}{c|}{Keras}	& \multicolumn{2}{c|}{TF} & \multicolumn{2}{c|}{Theano}	& \multicolumn{2}{c|}{Torch}     & \multirow{2}{*}{P value}\\ \cline{2-11} 
                              & \rotatebox[origin=c]{90} {SO}       & \rotatebox[origin=c]{90} {GitHub}       & \rotatebox[origin=c]{90} {SO}       & \rotatebox[origin=c]{90} {GitHub}        & \rotatebox[origin=c]{90} {SO}       & \rotatebox[origin=c]{90} {GitHub}        & \rotatebox[origin=c]{90} {SO}       & \rotatebox[origin=c]{90} {GitHub}       & \rotatebox[origin=c]{90} {SO}       & \rotatebox[origin=c]{90} {GitHub} &        \\ \hline
                    
Bad Performance         	& 31\% & 19\%		& 16\% & 14\%		& 8\%  & 8\%      	& 11\%    & 6\%		& 8\% & 24\%     	      		     	  &  0.9152    \\ \hline
Crash                   		& 40\% & 69\%		& 61\% & 86\%		& 77\% & 92\%     	& 70\%    & 20\%		& 60\%   & 16\%      	      		     	  & 0.7812    \\ \hline
Data Corruption         		& 6\%  & 4\%		& 5\%  & 0\%		& 6\%  & 0\%      	& 4\%     & 6\%			& 4\%   & 16\%     	      		      	  & 0.948    \\ \hline
Hang                    		& 0\%  & 0\%		& 0\%  & 0\%		& 1\%  & 0\%      	& 0\%     & 0\%			& 0\%      & 0\%      	      		      	   & 0.3466   \\ \hline
Incorrect Functionality 	& 23\% & 8\%		& 13\% & 0\%		& 7\%  & 0\%      	& 11\%    & 59\%		& 16\%     & 42\%          		      	  & 0.5418    \\ \hline
Memory Out of bound     	& 0\%  & 0\%		& 3\%  & 0\%		& 1\%  & 0\%      	& 4\%     & 0\%			& 0\%      & 0\%      	      		      	    & 0.0844  \\ \hline
Unknown                 		& 0\%  & 0\%		& 2\%  & 0\%		& 0\%  & 0\%     	& 0\%     & 9\%		& 12\%     & 2\%     	     		      	    & 0.8419  \\ \hline
\end{tabular}
\end{table}
	\section{Difficult Deep Learning stages}
In this section, we answer \textbf{RQ4} by studying the bugs happening at the different stage of the deep learning pipeline. We use the categorization of the posts about deep learning stages to analyze \textbf{RQ4}. 
\begin{figure}
	\includegraphics[keepaspectratio=true,scale=0.15]{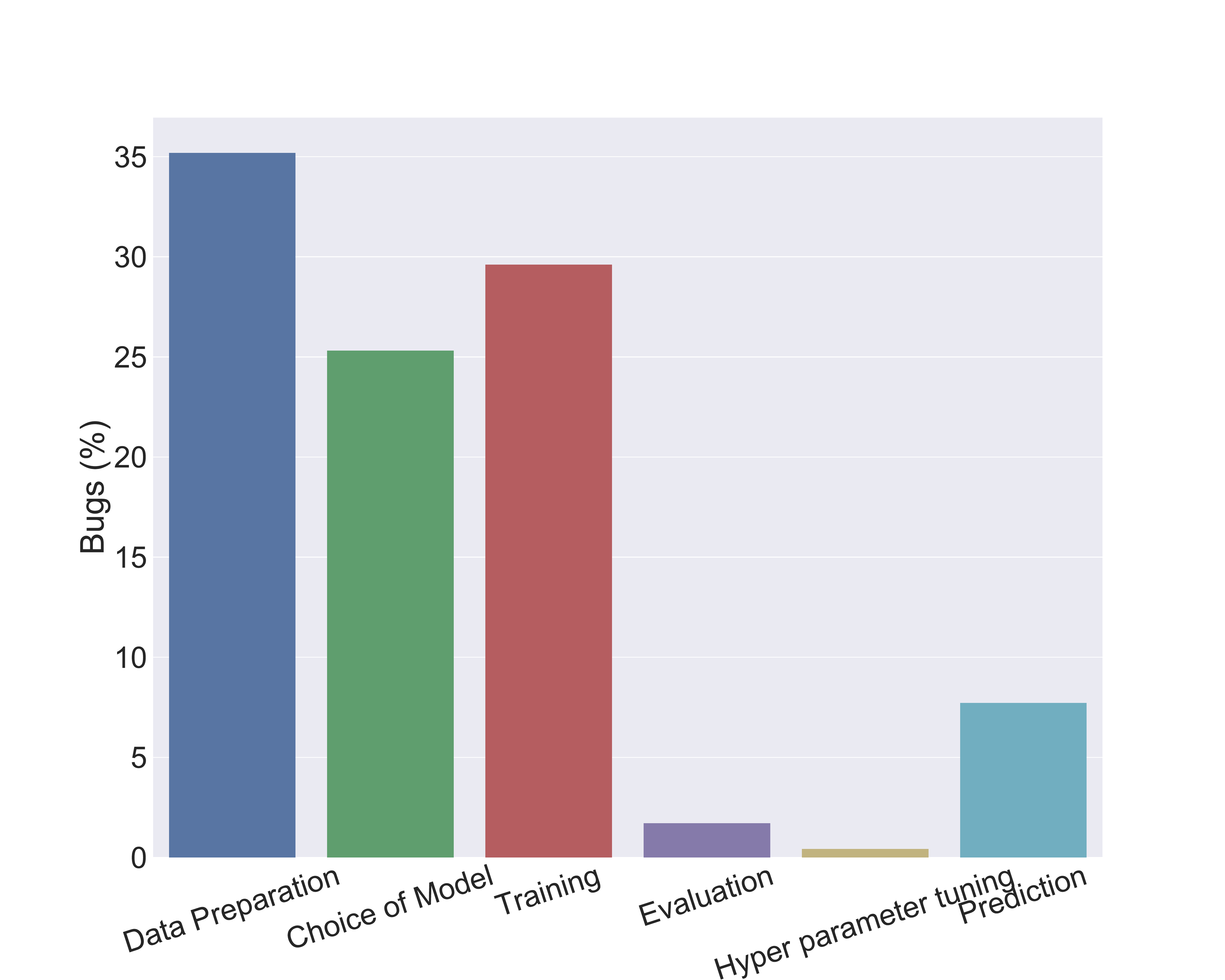}
	\caption{Bugs at different stages of the Deep Learning pipeline}
	\label{fig:stage}
\end{figure}
\subsection{Data Preparation }
\finding{32\% of the bugs are in the data preparation stage of the deep learning pipeline}
From \fignref{fig:stage} we see, most of the bugs in deep learning programming happen at the data preparation stage. 
\subsection{Training stage}
\finding{27\% of the bugs are seen during the training stage}
The next stage where most bugs happen is the Training stage which is kind of expected. A lot of bugs related to IPS and SI are from the training stage.

\subsection{Choice of model}
\finding{Choice of model stage shows 23\% of the bugs}
Choice of model is the third stage in terms of the likelihood to have bugs. 
In choice of model stage, we construct the model and chose the right algorithm. Major root causes of bugs in this stage are IPS, SI, and UT. 


\section{Commonality of Bug}

\begin{figure}
	\includegraphics[keepaspectratio = True, scale = .13]{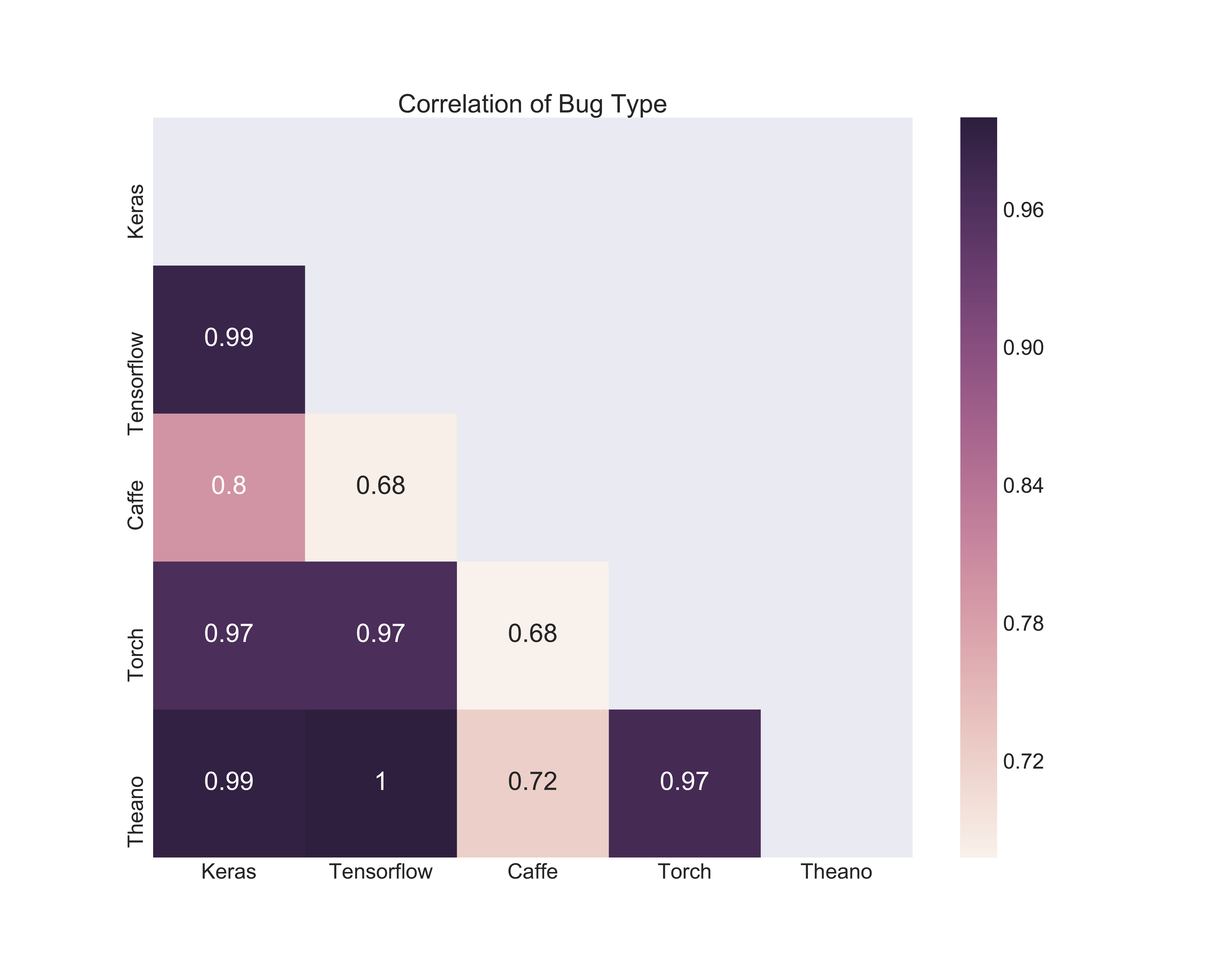}
	\vspace{-2.8em}
	\caption{Correlation of Bug Types among the libraries}
	\label{fig:btypecorr}
\end{figure}

In this section, we try to explore the answer to \textbf{RQ5} to identify whether
there is any relationship among the bugs in different deep learning libraries. Our primary
hypothesis was that the libraries will be strongly correlated based on the
distribution of bugs as they are doing the similar tasks. 

Our analysis confirms that hypothesis as shown in \fignref{fig:btypecorr}. 
We see that the libraries have a strong correlation coefficient close to 1. Surprisingly
\torch has shown very weak correlation with other libraries in terms of bug
type.
We then randomly studied the 30 posts having codes for each of the libraries to see whether we notice any
common antipatterns that can lead to this strong correlation of bug type. 

\begin{figure}
	\includegraphics[keepaspectratio = True, scale = .3]{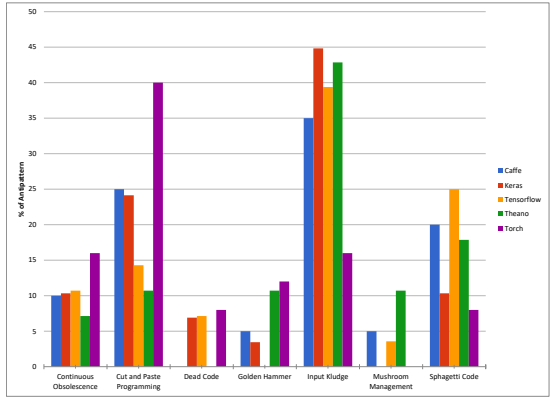}
	\vspace{-2.8em}
	\caption{Distribution of different antipatterns}
	\label{fig:antipat}
\end{figure}

\begin{figure*}
	\centering
	\begin{tabular}[b]{c}
		\includegraphics[keepaspectratio = True, scale= .25]{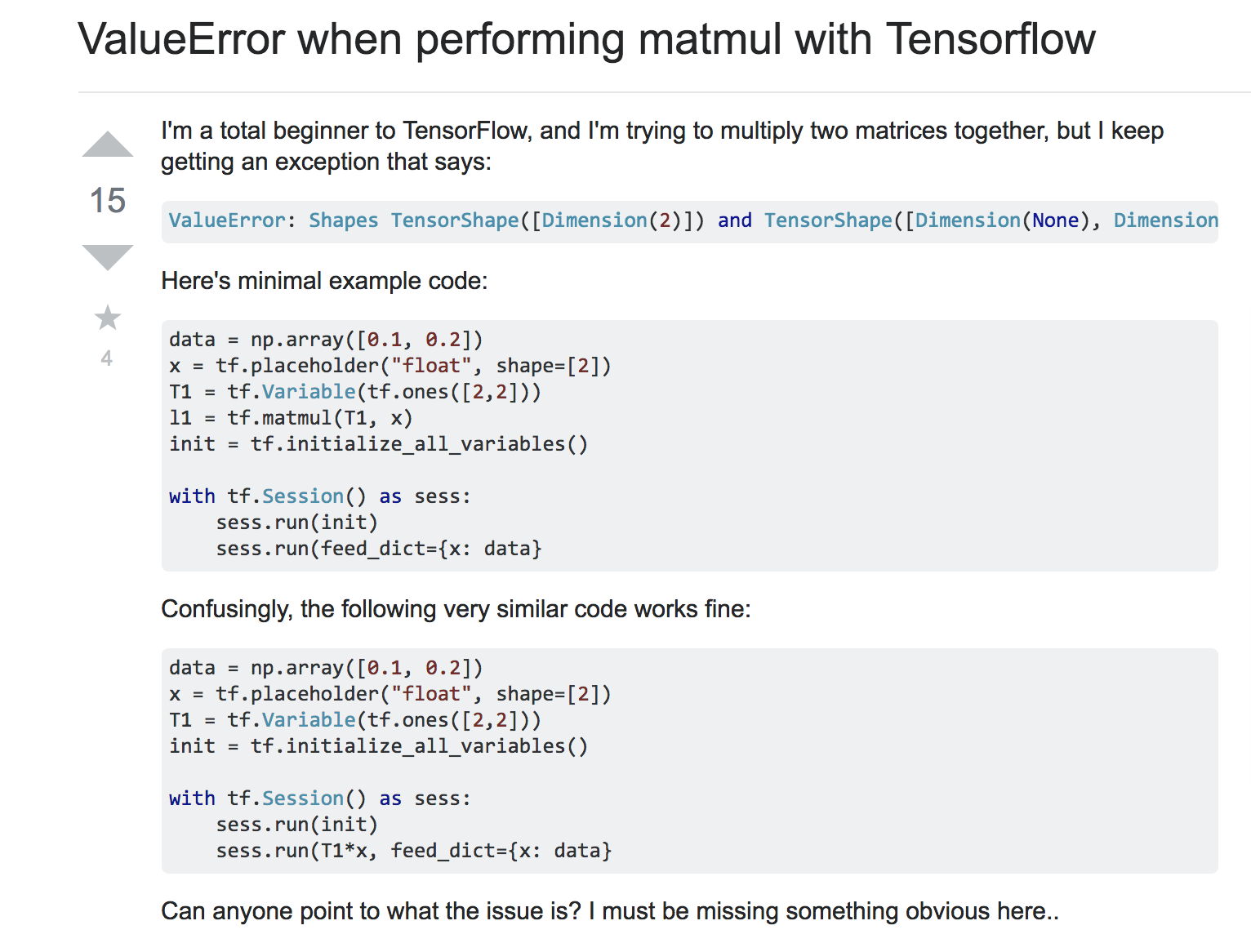} \\
		\small (a) Tensorflow Example of Input Kludge
	\end{tabular} \qquad
	\begin{tabular}[b]{c}
		\includegraphics[keepaspectratio = True, scale= .3]{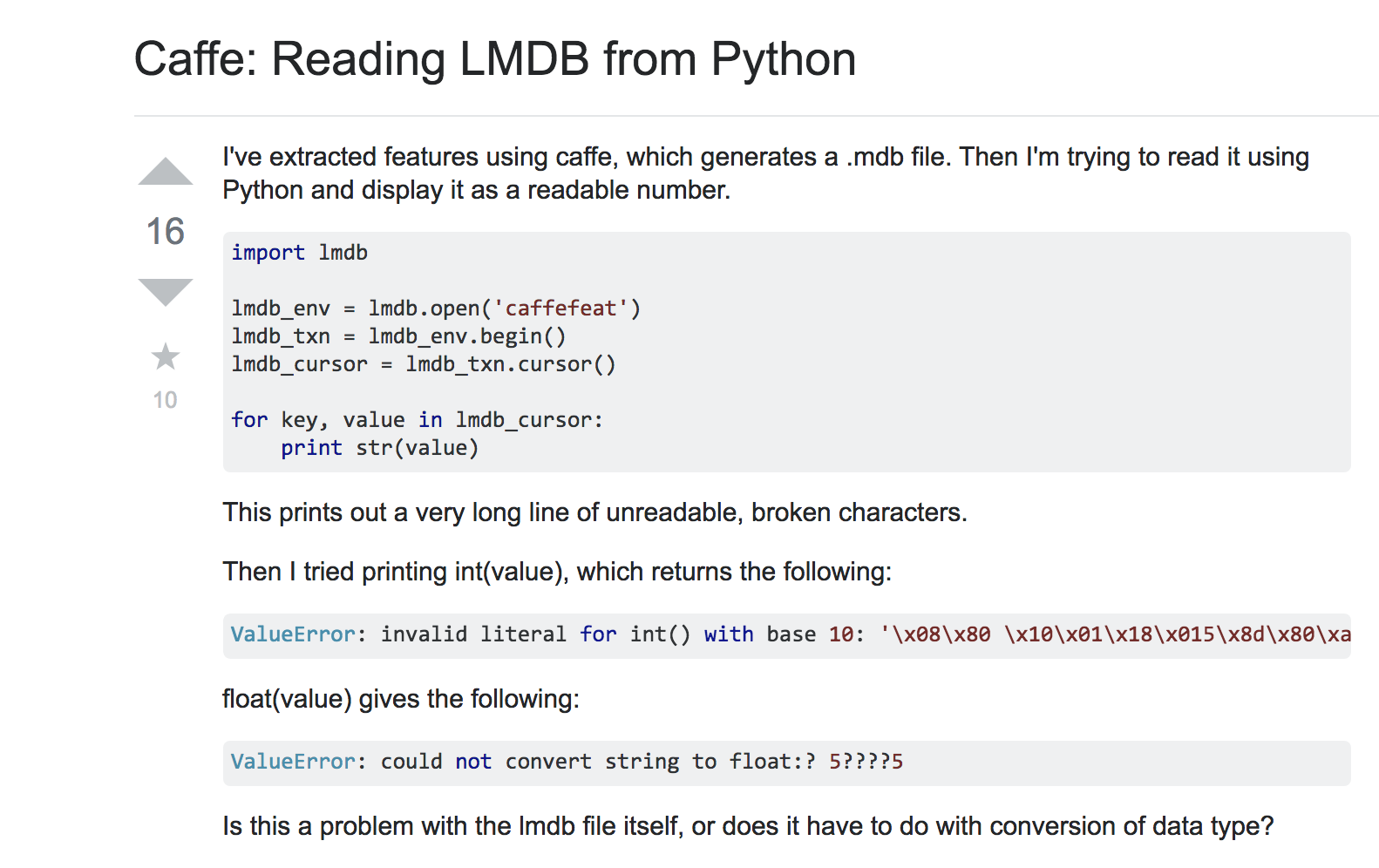} \\
		\small (b) Caffe Example of Input Kludge
	\end{tabular}
	\caption{Example of similar antipattern in \tensor and \caffe}
	\label{fig:antcomp}
\end{figure*}

\finding{\tensor and \caffe have a similar distribution of antipatterns while
	\torch has different distributions of antipatterns}
We have identified the antipatterns through deeper analysis
of the \sof buggy codes for further investigating the strong correlation of
\tensor and \caffe as well as the weak correlation of \torch and \caffe. The antipatterns found are
\textbf{Continuous Obsolescence, Cut-and-Paste Programming, Dead Code, Golden
	Hammer, Input Kludge, Mushroom Management, Spaghetti Code}. This classification is taken from \cite{antipatterns}.
The distribution of different antipatterns across the libraries is shown in
\fignref{fig:antipat}.
We see that in \tensor and \caffe 30\%+ of the antipatterns are Input Kludge. 
On the other hand, in \torch 40\% of the bugs happen due to the Cut-and-Paste 
Programming antipattern. This shows that the strong correlation
between the distribution of bugs in \torch and \caffe can be explained from the
similarity of common antipatterns for these two libraries. The weak
correlation between the distribution of \torch and \caffe bugs can be the result
of a dissimilar distribution of antipatterns between these two libraries. 
For example, we see \sof code snippets of Input Kludge antipatterns from \tensor and \caffe in the
example shown in \fignref{fig:antcomp}. Both of these programs can be easily broken by user input and the program does not perform sanity check on the inputs.

	\section{Evolution of Bugs}
In this section, we explore the answer to \textbf{RQ6} to understand how the bug patterns have changed over time. 

\subsection{Positive growth of Structural Logic Bugs}
\finding{In \keras, \caffe, \tensor Structural logic bugs are showing increasing trend}
From 2015 - 2018 Structural logic bugs in \caffe are respectively 30\%, 32\%, 67\%, 100\% indicating structural logic bugs are being discussed more by the developers since 2015. It is expected as deep learning started gaining increasing attention since 2015 and more developers started to use deep learning libraries to write software. 
\begin{figure}
	\includegraphics[keepaspectratio = True, scale= .2]{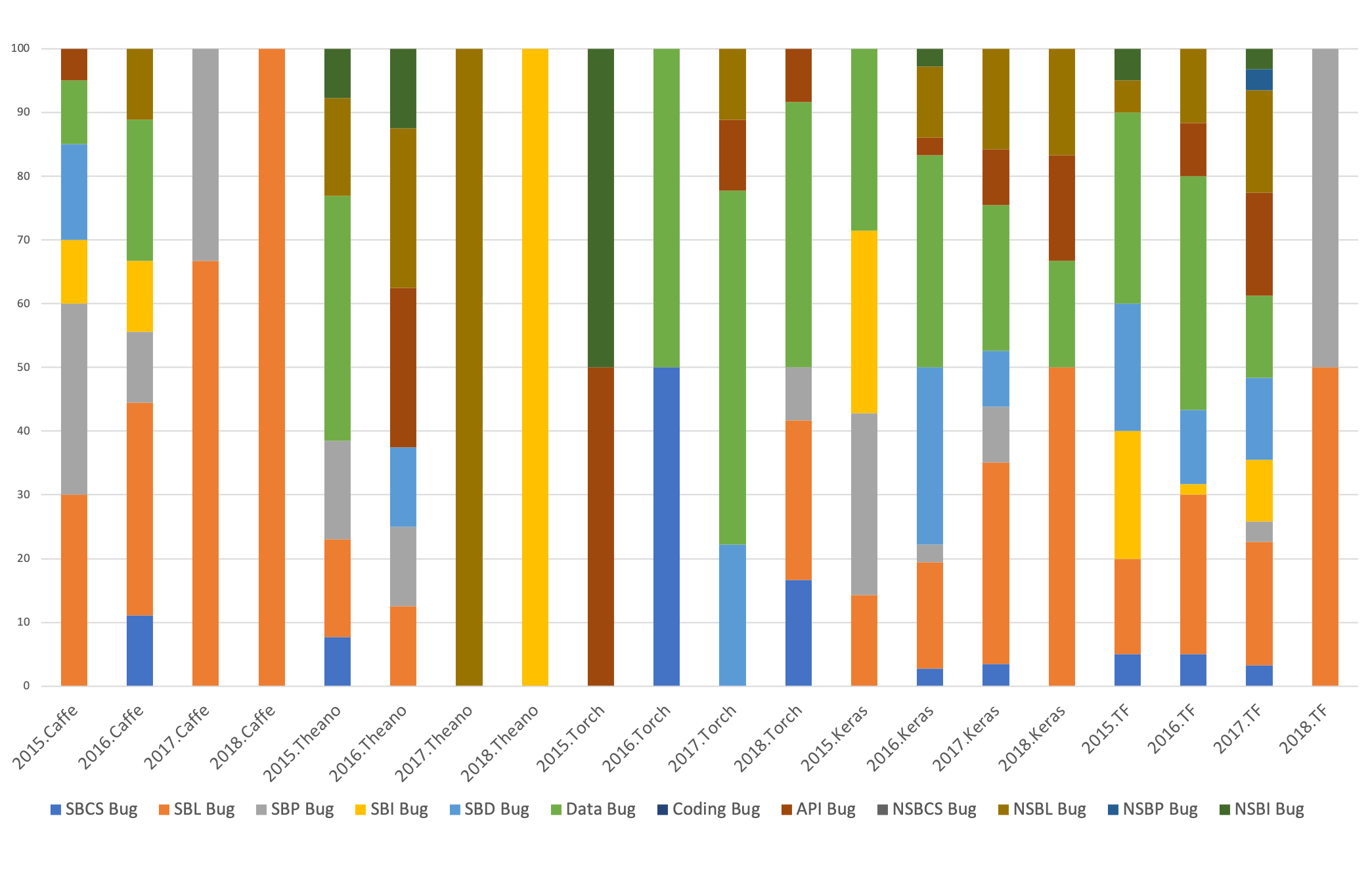}
	\caption{Timeline of Evolution of Bugs}
	\label{fig:timeline}
\end{figure}
\subsection{Decreasing trend of Data Bugs}
\finding{Data Bugs slowly decreased since 2015 except \torch}
In \torch Data Bugs stayed almost consistent maintaining close to 50\% of the bugs in discussed in 2016-2018. 
In \keras Data Bugs slowly decreased from 27\% - 15\% since 2015. 
In \tensor Data Bugs slowly decreased from 30\% - 10\% since 2015 - 2018. 
In the other two libraries also, the Data Bugs slowly decreased reaching close to 0. 
The possible reason for this trend is the development of popular specialized data libraries like \texttt{pandas} that enable exploratory data analysis to understand the properties of data better. Besides, the use of Tensor data type having type and shape information helps to get rid of some of the Data Bugs. Still more verification support in these libraries will help to get rid of these bugs. 
	\section{Threats to Validity}
\label{sec:tv}

{\bf Internal threat.\ }
One internal threat to the validity of our results could be 
our classification of the bugs.
We used the classification scheme from a vetted 
taxonomy~\cite{zhang2018empirical,beizer1984software} to 
classify the bugs. 
We also followed open coding scheme to add more types if needed. 
One PhD student was initially dedicated to go over all the posts 
to come up with additional classification scheme, if necessary. 
This whole process was monitored using pilot study. 
Another possible source of the threat is that the labeling 
of the data can be biased. 
To mitigate this threat two trained Ph.D. students independently 
studied the misuse posts to label them.
The inter-rater agreements  
was measured using Cohen's 
Kappa coefficient and the disagreements were reconciled under 
the monitoring of an expert. 
We conducted pilot study to continuously monitor the labeling 
process and conducted further training at 5\% and 10\% of the 
labeling where the Kappa coefficient was close to 0\% and 80\%. 

{\bf External threat.\ }
An external threat can be the trustworthiness of the 
dataset we collected. 
To avoid low-quality posts we only collected the posts that 
have score of at least 5. 
A score of 5 can be a good metric to trust the post as a good 
discussion topic among the programmer community that cannot 
merely be solved using some Google search. 
The reputation of the users asking question about deep learning 
can be another reason to question the quality of the posts. 
To alleviate this threat we have only studied top scored posts 
which are from users with different range of reputations (1 - 150K+). 
This indicates that the posts are from users ranging from newbie to experts. 
	\section{Discussion}
\label{sec:dis}


We have seen in the analysis of RQ1 that most of the bugs in deep learning
programming are Data Bugs. These type of Bugs can have drastic effect causing
the program to crash as well as leading to bad performance. In general, we see
the programmers have very limited or no access to data verification tools. It is
often confusing whether the data is in right format needed by the model, whether
the variables are properly encoded or not, whether there are missing data that
can cause the model to fail, whether the train test split is good enough,
whether the data is shuffled properly to avoid training bias etc. This finding
suggests that development of \textbf{data verification tools} can help
programmers solve a large number of data bugs. As deep learning models are
strongly coupled with data, \textbf{model analysis} tool to explore whether a
particular model is the right fit for the data in hand can help to resolve these
strong coupling of data and model related problems. 

We have also seen while
exploring RQ1 that structural logic bugs are the second major type of bugs. This
happens due to wrong logical organization of the model, hidden layers, using wrong
codes, etc. These kind of problems can be solved by some \textbf{automated model
and parameter recommendation} tools. How to develop these kind of tools need
further research. A methodology could be to mine large scale open source code
repositories and identify the common code patterns
and suggest examples from common code patterns used in qualified code bases.

	\section{Related works}
\label{sec:related}
\textbf{Empirical Study on Bugs}.
Thung \etal \cite{thung2012empirical} studied three machine learning systems, Apache Mahout, Lucene, and OpenNLP and manually categorize the bugs into different categories. They focused on bug frequencies, bug types, severity of the bug, bug-fixing duration, bug-fixing effort, and bug impact. Different from them, we focus on bug types, bug root causes, and bug impact of five deep learning libraries which are Tensorflow, Keras, Torch, Caffe, and Theano. 

Zhang \etal \cite{zhang2018empirical} investigated bugs from deep learning applications built on top of Tensorflow. They collected bugs from Stack Overflow questions and Github commits. Then, they manually studied the collected bugs based on three perspectives which are root cause, bug type, and impact. 
While Zhang \etal have certainly charted the course,  
we have studied a cross-section of five deep learning libraries with different design 
constraints.
Furthermore, besides root cause, bug type, and impact, we also focus on bugs in deep learning stages which includes data preparation, modeling, training, evaluation, tuning, prediction. We have also studied some common antipatterns to explain the strong correlation of bugs in these libraries as bugs have strong relation to antipatterns \cite{taba2013predicting}.

There are some empirical studies focused on specific types of bugs. Lu {\em et. al.} \cite{lu2008learning} studied real-world concurrency bug characteristics. Gao {\em et. al.} \cite{gao2018empirical} conducted an empirical study on recovery bugs in large-scale distributed systems. API changes problems was studied by \cite{bavota2015impact, dig2006apis,li2013does,kula2018empirical}. Our work focuses on the bugs in the usage of deep learning libraries.

\textbf{Bugs classification}. %
Classification of API-misuse from different perspectives and domains have 
been done previously. Classification of software defects by IEEE served as 
the basis for IBM's orthogonal defect classification (ODC) as discussed in 
\cite{chillarege1992orthogonal}. The defect types include conceptual program 
elements which are \emph{function}, \emph{check}, \emph{documentation}, 
\emph{assignment}, \emph{algorithm} and \emph{violation type}. More recently, 
\cite{xu2016python} presents a Python predictive analysis. Their tool is able to detect 46 bugs with 16 unreported before
and they also classify those bugs base on their defect in the Python program. Our study focus on deep learning bugs, we classify the bugs based on bug type, root cause, and effect.

\textbf{\sof and \gh study}. %
Meldrum {\em et. al.}~\cite{meldrum2017crowdsourced} studied 266 papers 
using \sof platforms to show the growing impact of \sof on software engineering research. 
Kavaler \cite{kavaler2013using,linares2014api} used \sof to analyze Android APIs. 
Barua {\em et. al.} \cite{barua2014developers} analyzed the
textual content of \sof discussions to understand the thoughts and needs of developers. 
These works have not studied bugs in deep learning software.
%

	\section{Conclusion}
\label{sec:conclusion}
Although deep learning has gained much popularity and strong developer community in recent years, developing software using existing deep learning libraries can be error-prone.
In this paper, we have presented an empirical study to explore the bugs 
in software using deep learning libraries. 
In our study we have studied \texttt{2716} qualified \sof posts and \texttt{500} \gh bug fix commits to identify the bug types, root causes of bugs, effects of bugs in usage of deep learning. We have also performed an inter-stage analysis to identify the stages of deep learning pipeline that are more vulnerable to bugs.  We have also studied the buggy codes in \sof to find antipatterns leading to bugs to understand the strong correlation of the bug types in deep learning libraries. 
Our study found that data bug and logic bug are the most severe bug types 
in deep learning software appearing more than 50\% of the times.
Major root causes of these bugs are Incorrect Model Parameter (IPS) and Structural Inefficiency (SI).
Last but not least, bugs in the usage of deep learning libraries 
are strongly correlated. 
	\bibliographystyle{ACM-Reference-Format}
	\bibliography{refs}
	
\end{document}